\documentclass[twocolumn,aps,prc,showpacs]{revtex4}
\usepackage{amsmath,bm}
\usepackage{graphicx}

\begin{document}

\draft
\title{Surveying the Nucleon-Nucleon Momentum Correlation Function
in the Framework of Quantum Molecular Dynamics Model}
\author{Y. G. Ma} \thanks{Email: ygma@sinap.ac.cn}
  \affiliation{Shanghai Institute of Applied Physics, Chinese Academy of Sciences, P.O. Box 800-204,
Shanghai 201800, China}

\author{ Y. B. Wei}
\affiliation{Shanghai Institute of Applied Physics, Chinese
Academy of Sciences, P.O. Box 800-204, Shanghai 201800, China}
\author{ W. Q. Shen}
\author{X. Z. Cai}
\author{J. G. Chen}
\author{J. H. Chen}
\author{D. Q. Fang}
\author{W. Guo}
\author{C. W. Ma}
\author{G. L. Ma}
\author{Q. M. Su}
\author{W. D. Tian}
\author{K. Wang}
\author{T. Z. Yan}
\author{C. Zhong}
\author{J. X. Zuo}

\affiliation{Shanghai Institute of Applied Physics,
 Chinese Academy of Sciences, P.O. Box 800-204,
Shanghai 201800, China}

 \date{\today}

\begin{abstract}

Momentum correlation functions  of the nucleon-nucleon pairs are
presented for reactions with C isotopes bombarding a $^{12} \rm C$
target within the framework of the isospin-dependent quantum
molecular dynamics model. The binding-energy dependence of the
momentum  correlation functions is also explored, and other
factors that have an influence on momentum correlation functions
are investigated. These factors include momentum-dependent nuclear
equation of state, in-medium nucleon-nucleon cross sections,
impact parameters, total pair momenta, and  beam energy. In
particular, the rise and the fall of the strength of momentum
correlation functions at lower relative momentum are shown with an
increase in beam energy.

\end{abstract} ¡¡¡¡

 \maketitle

\section{ Introduction}

The method of two-particle intensity interferometry was developed
by Hanbury-Brown and Twiss \cite{Brown} in the early 1950s.
Originally they applied two photon correlation to measure the
angular diameter of stars and other astronomical objects.
Initially, the method did not receive universal acceptance before
a number of terrestrial experiments were performed to confirm it.
Now the method of intensity interferometry is commonly referred as
the Hanbury-Brown/Twiss (HBT) effect. Although the original
application of the HBT effect used photons as the detected
particles, it was rapidly realized that the approach can be
generalized to include correlation measurements for other bosons
and fermions as well. The first measurements of the HBT effect in
subatomic physics came from elementary-particle reactions.
Goldhaber {\it et al.} extracted the spatial extent of the
annihilation fireball in proton-antiproton reactions from two-pion
correlations \cite{Goldhaber}. In fact, the method explores that
identical particles situated nearby in phase-space experience
quantum statistical effects resulting from the
(anti)symmetrization of the multi-particle wave function. For
Bosons, therefore, the two-particle coincidence rate shows an
enhancement at small momentum difference between the particles.
The momentum range of this enhancement can be related to the size
of the particle source in coordinate space. Recently, there are
not only substantial experimental literatures on the technical
applications, but also a large number of theoretical papers on
momentum correlations with different models from low energy to
relativistic energy heavy ion collisions. For reviews, see the
references \cite{Boal,Bauer,Heinz2,Wiedman}. More recently, HBT
method has been extended to others fields, for instance, the
analogous correlations in semiconductors and in free space aiming
at the fermionic statistics of electrons \cite{Henny,Picciotto}.

Many experimental measurements of HBT effect have been performed
for heavy ion collisions (HIC) at intermediate energy in recent
years. With the increasing of beam energies,  nucleon-nucleon
collision plays a  dominate role during the reaction process in
the intermediate energy, which results in an increasing importance
of earlier particle emission. By the application of the
two-particle correlation function, one could obtain the
information on particle emission  and collision dynamics.

In this energy domain, most correlation measurements focus on
two-proton correlation functions. The shape of a two-proton
correlation function reflects the combined effects of the Pauli
blocking principle, Coulomb interaction  and proton-proton ($p-p$)
nuclear interaction. Earlier measurements of two-proton
correlation functions in intermediate energy heavy ion collisions
performed by Lynch  {\it et\ al.}  \cite{Lynch} provided the
evidence for particle emission from localized highly-excited
regions. After that, many experimental groups have investigated
the momentum correlation functions in various aspects, such as for
the unstable particle populations \cite{Pochodzalla}, the
dependence on the impact parameter \cite{Gong}, the dependence of
the total momentum of nucleon pairs \cite{Colonna}, the dependence
of the isospin of the emitting source \cite{Ghetti} and  so on.
More interestingly, HBT technique has been used to construct
neutron-neutron correlation function which is useful to
investigate the properties of neutron-halo nuclei
\cite{Orr,Marques1,Marques2}. The details about the nuclear
Equation of State (EOS) and the collision dynamics could be
revealed from the correlation function  by the comparison between
the experimental data and the transport model calculations.

However, in most studies of $p-p$ correlation functions, the HBT
strength at 20 MeV/{\it c} of the proton-proton relative momentum
is taken a unique quantity to determine the source size and/or
emission time of two proton emission. In a recent analysis for the
HBT data below 100 MeV/nucleon with the imaging method, Verde {\it
et\ al.}  show that the width of  correlation function provides
the information of the source size of the fast dynamical component
while the peak of the correlation function is sensitive to
relative yield from slow and fast emission components
\cite{Verde1,Verde2}. In addition, it was claimed that the proton
emission from slow statistical component was not suitably treated
in the conventional transport model, such as BUU model
\cite{Buu1}. In light of the above studies, the whole shape of the
correlation function is important for deducing space-time
information  of the emission source \cite{Verde1}. For heavy ion
reactions with mid-heavy projectile and target combination in the
Fermi energy domain, the slow emission component should not be
neglected. To minimize the complication of the slow and fast
component of proton emission in correlation function, it might be
useful to choose light reaction systems at higher beam energies.
In this context, we will use  C + $^{12}$C system to investigate
momentum correlation functions above 100 MeV/nucleon in this work.

To understand the details of collisions for different reactions by
the HBT studies, a reliable simulation of the collision dynamical
process for the heavy ion reaction is required. The simulation
shall give a reasonable treatment of fragment formation after the
final state interaction. There are some good event-generator
models to describe the collision process. In relativistic heavy
ion collisions, successful models include both the
string-hadronic-like models, such as relativistic quantum
molecular dynamics and the parton cascade model etc. In
intermediate energy region, the successful transport model
includes BUU model \cite{Bertsch} and  Quantum Molecular Dynamics
(QMD) model \cite{Aichelin}. From those event-generator models one
can obtain the phase space of the emitted particles with different
parameters of the EOS and then construct the momentum correlation
function. Recently, the nuclear symmetry energy dependence of HBT
has been also explored through the isospin dependent BUU model
\cite{Chen1} in intermediate energy heavy ion collisions.
Moreover, the nuclear binding energy and separation energy
dependences of the HBT strength have been investigated with the
help of the Isospin Dependent Quantum Molecular Dynamics model
(IDQMD) \cite{Wei1,Wei2}. In this paper, more features of momentum
correlation function are reported by using the IDQMD model.

The paper is organized as follows: in Sec. II we describe the HBT
technique and the IDQMD model. The stability of the IDQMD model is
checked; Section III presents the results and discussions. We
discuss the influences of the following ingredients:
initialization of projectile and target, emission time of
nucleons, evolution time of the reaction, gate on the total
momentum of nucleon-nucleon pair, soft and stiff momentum
dependent EOS, in-medium nucleon-nucleon cross section, impact
parameter and incident energy etc. Finally the conclusions are
presented in Section IV.

\section{HBT Technique and the IDQMD model}

Firstly, we recall the HBT technique.¡¡As we know, the wave
function of relative motion of light identical particles is
modified by the final-state interaction  and quantum statistical
symmetries when they are emitted in close proximity in space-time,
and this is the principle of intensity interferometry, i.e. HBT.
In standard Koonin-Pratt formalism \cite{Koonin,Pratt1,Pratt2},
the two-particle correlation function is obtained by convoluting
the emission function $g(\mathbf{p},x)$, i.e., the probability for
emitting a particle with momentum $\mathbf{p}$ from the space-time
point $x=(\mathbf{r},t)$, with the relative wave function of  two
particles,
\begin{equation}
C(\mathbf{P},\mathbf{q})=\frac{\int d^{4}x_{1}d^{4}x_{2}g(\mathbf{P}%
/2,x_{1})g(\mathbf{P}/2,x_{2})\left| \phi (\mathbf{q},\mathbf{r})\right| ^{2}
}{\int d^{4}x_{1}g(\mathbf{P}/2,x_{1})\int
d^{4}x_{2}g(\mathbf{P}/2,x_{2})}, \label{CF}
\end{equation}%
where  $\mathbf{P(=\mathbf{p}_{1}+\mathbf{p}_{2})}$ and $\mathbf{q(=}%
\frac{1}{2}(\mathbf{\mathbf{p}_{1}-\mathbf{p}_{2}))}$ are the
total and relative momenta of the particle pair respectively,
 and $\phi (\mathbf{q}, \mathbf{r})$ is the relative two-particle
 wave function with
their relative position $\mathbf{r=(r}_{2}\mathbf{-r}_{1}%
\mathbf{)-}$ $\frac{1}{2}(\mathbf{\mathbf{v}_{1}+\mathbf{v}_{2})(}t_{2}-t_{1}%
\mathbf{)}$. This approach is very useful in studying effects of
nuclear equation of state and nucleon-nucleon cross sections on
the reaction dynamics of intermediate energy heavy-ion collisions
\cite{Bauer}.

In the viewpoint of theoretical simulation, the correlation
function can be established by using an event generator that
produces the phase space information by modelling the collision
dynamics and particle production. The event-generator correlation
functions are then constructed from the positions and momenta
representing the single-particle emission distribution at the time
of the last strong interaction, i.e. at freeze-out. In the present
work, the event-generator  is the Isospin-dependent Quantum
Molecular Dynamics transport model \cite{Aichelin}, which has been
successfully applied to the HBT studies of the heavy-ion
collisions for neutron-rich nuclei induced reactions
\cite{Wei1,Wei2}. Using a computation code named CRAB (
Correlation After Burner )
 of Pratt \cite{crab}, which takes into account final-state
nucleon-nucleon interactions, we have evaluated two-nucleon
correlation functions from the emission function given by the
IDQMD model. In the following part, we shall introduce the model
briefly.

The Quantum Molecular Dynamics approach  is a many body theory to
describe heavy ion reactions from intermediate energy to 2
GeV/nucleon \cite{Aich2}. It includes several important
ingredients: initialization of the target and the projectile;
nucleon propagation in the effective potential; nucleon-nucleon
collisions in nuclear medium; Pauli blocking effect and the
numerical test. A general review about the QMD model can be found
in \cite{Aichelin}.  The IDQMD model is based on the QMD model
affiliating the isospin factors.

The heavy-ion collision dynamics at intermediate energies is
mainly governed by three components: the mean field, two-body
collisions and Pauli blocking. Therefore, for an isospin-dependent
reaction dynamics model it is important to affiliate  isospin
degrees of freedom with the above three components. In addition,
the sampling of phase space of neutrons and protons in the
initialization should be treated separately because of larger
difference between neutron and proton density distributions for
nuclei far from the $\beta$-stability line. For exotic
neutron-rich nucleus one should sample a stable initialized
nucleus with neutron-skin or halo structure so that one can
directly incorporate nuclear structure effects into a microscopic
transport process. The IDQMD model has been improved based on the
above ideas and the details will be given in the following.

In the present calculations the interaction potential in the IDQMD
is determined as follows:
\begin{equation}
U(\rho ) = U^{\rm Sky} + V^{\rm Coul} + U^{\rm sym} + V^{\rm Yuk}
+ U^{\rm MDI} + U^{\rm Pauli},
\end{equation}
where $U^{\rm Sky}$ is the density-dependent Skyrme potential and
it reads when the momentum dependent potential is included
\begin{equation}
U^{\rm Sky} = \alpha (\frac \rho {\rho _0}) + \beta (\frac \rho
{\rho _0})^\gamma + t_4 {\rm ln}^2[\varepsilon(\frac \rho {\rho
_0})^{2/3}+1]\frac \rho {\rho _0},
\end{equation}
where $\rho$ and $\rho_0$ are total nucleon density and its normal
value, respectively. The parameters $\alpha$, $\beta$, $\gamma$,
$t_4$ and $\varepsilon$ are related to the nuclear equation of
state and listed in Table. 1. $V_c$ is Coulomb potential. $U^{\rm
Yuk}$ is the Yukawa potential ,
\begin{equation}
U^{\rm Yuk} = t_3 \frac{ exp(\frac{\left|\overrightarrow{r_1}-
\overrightarrow{r_2}\right|}{m})}{
\frac{\left|\overrightarrow{r_1}- \overrightarrow{r_2}\right|}{m}}
\end{equation}
where $m$ = 0.8 fm.
 $U^{\rm MDI}$ is the momentum dependent interaction
\cite{Aich2},
\begin{equation}
U^{\rm MDI} = t_4{\rm
ln}^2[t_5(\overrightarrow{p_1}-\overrightarrow{p_2})^2+1]\frac
\rho {\rho _0}
\end{equation}
where $\overrightarrow{p_1}$ and $\overrightarrow{p_2}$ are the
momentum of two interacting nucleons.
$U^{\rm Pauli}$ is the Pauli
potential ,
\begin{equation}
U^{\rm Pauli} = V_p\ (\frac{\hbar} {p_0 q_0})^3 exp(-\frac{(\overrightarrow{r_i}-%
\overrightarrow{r_j})^2}{2q_0^2}-\frac{(\overrightarrow{p_i}-\overrightarrow{%
p_j})^2}{2p_0^2} )\delta _{p_i p_j}
\end{equation}
where
\[
\delta _{p_i p_j} = \left\{
\begin{array}{ll}
1 & \mbox{for neutron-neutron or proton-proton} \\
0 & \mbox{for neutron-proton}.
\end{array}
\right.
\]
The parameters $V_p$, $p_0$ and $q_0$ is 30 MeV, 400 MeV/{\it c}
and 5.64 fm, respectively. $U^{\rm sym}$ is the symmetry
potential. In the present calculation, we use $U^{\rm sym} =
C_{\rm sym}\frac{(\rho_{n} - \rho_{p})}{\rho_{0}}\tau_{z}$ where
$C_{\rm sym}$ is the strength of symmetry potential, taking the
value of 32 MeV. $\rho_n$ and $\rho_p$ are the neutron density and
the proton density, respectively. $\tau_{z}$ is the $z$th
component of the isospin degree of freedom, which equals  1 or -1
for neutrons or protons, respectively. The parameters of the
interaction potential are given in Table. 1 where $K$ = 200 or 380
MeV means the soft- or the stiff- momentum dependent potential,
respectively.

\begin{center}

\scriptsize

\begin{tabular}{|c|c|c|c|c|c|c|c|}\hline
$\alpha$ & $\beta$&$\gamma$&$t_{3}$&$t_{4}$&$t_{5}$ & $\varepsilon$ & $K$\\
\hline (\rm MeV)&(\rm MeV)&&(\rm MeV)&(\rm MeV)&($MeV^{-2}$)&(\rm
MeV)&(\rm MeV)\\\hline
-390.1&320.3&1.14&7.5&1.57&$5\times10^{-4}$&21.54&200\\\hline
-129.2 &59.4&2.09 &7.5&1.57&$5\times10^{-4}$&21.54&380\\\hline
\end{tabular}\\
\vskip 0.3 in Table 1. The parameters of the interaction
potentials
\end{center}

The in-medium nucleon-nucleon (NN) cross section can be
parameterized as isospin dependent from the available experimental
data. Studies of collective flow in HIC at intermediate energies
revealed the reduction of the in-medium NN cross sections
\cite{Westfall,Klakow,Ma_flow}. An empirical expression of the
in-medium NN cross section \cite{Klakow} is used:
\begin{equation}
\sigma_{\rm NN}^{\rm med} = (1 + f \frac{\rho}{\rho_{0}})
\sigma_{\rm NN}^{\rm free}
\end{equation}
with the factor $f$ $\approx$ - 0.2 which has been found to better
reproduce the flow data \cite{Westfall}. Here $\sigma_{\rm
NN}^{\rm free}$ is the experimental NN cross section \cite{Chen3}.
The free neutron-proton cross section $\sigma^{\rm free}_{\rm NN}$
is about a factor of 3 times larger than the free neutron-neutron
or proton-proton cross section below about 400 MeV/nucleon in the
laboratory energy. It should be mentioned that the relationship
between the neutron-proton cross section and neutron-neutron
(proton-proton) cross section depends also on the modification of
the nuclear density distributions during the reactions.

The Pauli blocking effect in IDQMD model is treated  separately
for  the neutron and the proton: Whenever a collision occurs,  we
assume that each nucleon occupies a six-dimensional sphere with a
volume of $\hbar^{3}$/2 in the phase space (considering the spin
degree of freedom), and then calculate the phase volume, $V$, of
the scattered nucleons being occupied by the rest nucleons with
the same isospin as that of the scattered ones. We then compare
2$V$/$\hbar^{3}$ with a random number and decide whether the
collision is blocked or not.

When the initialization of the projectile and the target is taken
in the IDQMD, the density distributions of protons and neutrons
are distinguished from each other. The references of neutron and
proton density distributions for the initial projectile and target
nuclei in IDQMD are taken from the Skyrme-Hartree-Fock (SHF)
method with parameter set SKM*. Using this density distribution,
we could get the initial coordinate of nucleons in nuclei in terms
of the Monte-Carlo sampling method. The momentum distribution of
nucleons is generated by means of the local Fermi gas
approximation:
 \begin{equation}
P^{i}_{F}(\vec r) = \hbar(3\pi^{2}\rho_{i}(\vec r))^{\frac{1}{3}},
(i= n,p).
\end{equation}
In the model, the radial density can be written as:
\begin{equation}
\rho(r) = \sum_{i}\frac{1}{(2\pi L)^{3/2}} exp(-\frac{r^{2} +
r_{i}^{2}}{2L})\frac{L}{2rr_{i}} \nonumber
\end{equation}

\begin{equation}
\times[exp(\frac{rr_{i}}{L}) - exp(-\frac{rr_{i}}{L})]
\end{equation}
where $L$ is the so-called Gaussian wave width (here $L$ = 2.16
${\rm fm}^2$).

 Stability  of the initialized nucleus has been checked by the
time evolution of the system at zero temperature \cite{Wei2}. The
accepted configurations are quite stable: only a few percentage
nucleons escape from the nucleus till 200 fm/${c}$ in the
intermediate energy domain. In addition, the stability is also
checked by tracking the time evolutions of the average binding
energy and the root mean square radius of the initialized nucleus
and good enough stability has been found. Lighter nuclei are
somewhat less stable. One or two out of ten nuclei lose a nucleon
in the required time span. To avoid taking  unstable
initialization of projectile and target in the IDQMD calculation,
we only select the initialization samples of nuclei which meet the
required stability.

Nuclear fragments are constructed by a modified isospin-dependent
coalescence model, in which particles with relative momentum
smaller than $p_{0}$ = 300 MeV/$c$ and relative distance smaller
than $R_{0}$ = 3.5 fm will be combined into a cluster.

In our calculations, the reactions of C-isotopes with $^{12}$C
have been performed. Most simulations are done in head-on
collisions ($b$ = 0 fm) at 100 MeV/nucleon or 800 MeV/nucleon. The
momentum correlation function  are constructed by the phase space
points at 200 fm/$c$ when the system is basically at the
freeze-out stage.

\section{Results and Discussions}

\subsection{ Stable initialization vs random initialization}

\begin{figure}
\vspace{-0.8truein}
\includegraphics[scale=0.4]{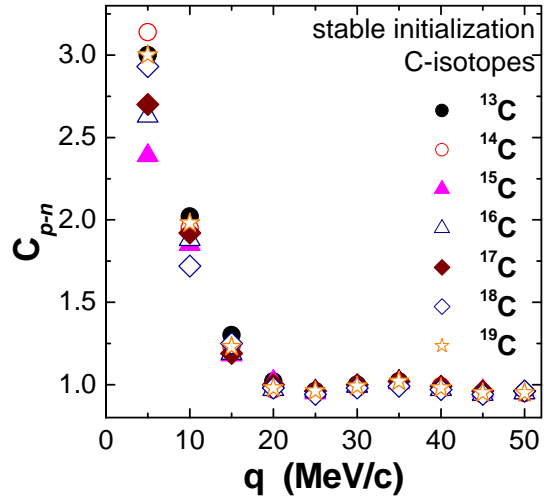}
\vspace{-0.5truein} \caption{ (Color online) The proton-neutron
correlation function  $C_{\it p-n}$ for the reactions of a chain
of C-isotopes with $^{12}$C at 800 MeV/nucleon  and $b$ = 0 fm
using the stable initialization of  projectile and target. The
meaning of symbols are illustrated in the insert. }
\label{Fig-HBT-C-stable}
\end{figure}

\begin{figure}
\vspace{-1.truein}
\includegraphics[scale=0.4]{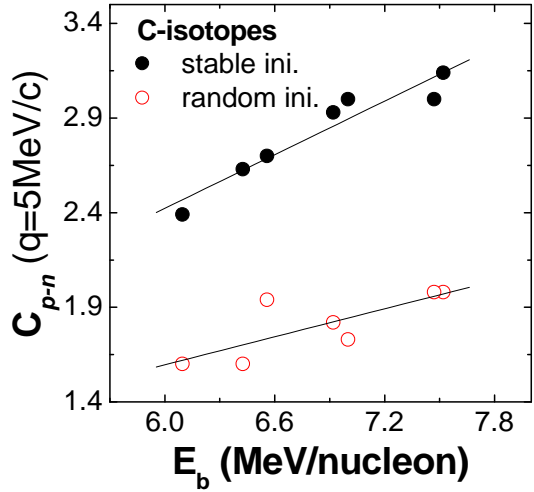}
\vspace{-0.5truein} \caption{ (Color online) The relationship
between the proton-neutron correlation function $C_{\it p-n}$ at 5
MeV/{\it c} and the average binding energy per nucleon of
C-isotopes. The filled circles represent the results with the
stable initial phase space taking in the IDQMD model  according to
the SHF density distribution. The open circles represent the
results with the random initial phase space. The collisions
 were simulated at 800 MeV/nucleon and $b$ = 0 fm. The target is $^{12}$C.
The lines are linear fits to guide the eyes.}
\label{Fig-HBTstrength-Eb}
\end{figure}

Fig.~\ref{Fig-HBT-C-stable} shows proton-neutron correlation
functions  from the reactions induced by a chain of C-isotopes
projectile on $^{12}$C target at 800 MeV/nucleon of  incident
energy and head-on collisions (impact parameter {\it b} = 0 fm)
when the suitable selection of the stable initialization is used.
In this figure the HBT strength of each isotope in lower relative
momentum can be separated. If we plot this strength at 5 MeV/{\it
c} ($C_{\it p-n}$) as a function of the mean binding energy
($E_b$) of the projectiles (C-isotopes), we find that there exists
an approximate linear relationship between $C_{\it p-n}$ and $E_b$
as it is shown by the solid circles in
Fig.~\ref{Fig-HBTstrength-Eb}.

In order to investigate the importance of initialization in the
HBT study, a comparison has been performed using a random
initialization for the projectile and the target in the IDQMD
based on Monte-Carlo sampling. In random initialization, the phase
space of the nucleons is generated with Monte Carlo random
sampling. During sampling of each nucleon, if the distance between
every two nucleons is larger than 1.5 fm  and  the product of the
space radius and the momentum radius between two nucleons is
larger than one constant according to the uncertain relationship,
the sampling will be accepted. In this case, there is no
additional requirement of the binding energy of the projectile or
the target.

\begin{figure}
\vspace{-.8truein}
\includegraphics[scale=0.4]{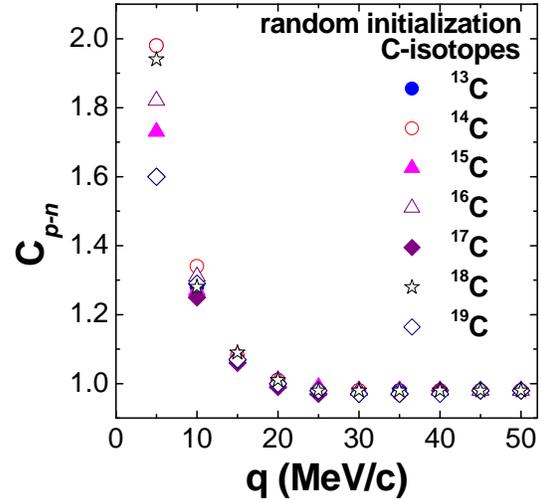}
\vspace{-0.5truein} \caption{ (Color online) The proton-neutron
correlation function  $C_{\it p-n}$ for the reactions of a chain
of C-isotopes with $^{12}$C target at 800 MeV/nucleon using the
random initialization of the projectile and the target. Symbols
which represent the different isotopes are illustrated in the
insert. } \label{Fig-HBT-C-rand}
\end{figure}

Taken such a random initialization, the initial phase space of the
nucleons for the projectile and the target is different event by
event. Through the transport process of IDQMD, we can obtain the
momentum correlation function and extract the strength of HBT in
the final states. Fig.~\ref{Fig-HBT-C-rand} shows the
neutron-proton HBT for different C-isotopes by random sampled
initialization. Similar to Fig.~\ref{Fig-HBT-C-stable}, the
strength at 5 MeV/{\it c} is not a constant for different isotopes
induced reactions which is shown by open circles in
Fig.~\ref{Fig-HBTstrength-Eb}. However, two apparent differences
can be observed between the results of stable initialization and
random initialization. One is the difference in the  magnitude of
the HBT strength. The values of  $C_{\it p-n}$ at 5 MeV/{\it c}
with the random initialization are less than those with the stable
initialization. For the random initialization, the initial phase
space  which may not meet the requirement of the ground state as
required in SHF calculation fluctuates event by event. In this
case the tightness of initial nucleus becomes weaker than that
with stable initialization of phase space which is sampled by the
SHF density distribution. Thus, the values of $C_{\it p-n}$ at 5
MeV/{\it c} which can embody the tightness between the nucleons
become smaller. The other difference is the slope of $C_{\it p-n}$
vs $E_b$. A steeper linear relationship is observed for the stable
initial phase space, while the dependence of $C_{\it p-n}$ vs
$E_b$ becomes weaker in the random case.

The comparison between both different initialization methods
indicates that the reasonable initial phase space of the
projectile which is sampled by the experimental $E_{b}$ and SHF
density calculation in the IDQMD is important and suitable to
investigate the dependence of the binding energy for some
observables. In the following calculations, we will use the stable
initialization to study some features of momentum correlation
functions.

\subsection{ Emission time of nucleons and evolution time of reaction}

In the intermediate energy domain, the extraction of the
space-time information is further complicated by two effects. One
is the presence of multiple sources of particle emission
\cite{Ghett_mult}, another is the different time scale of
statistical and dynamical emission from equilibrium and
non-equilibrium sources \cite{Verde1,Verde2}. The total
momentum-gated correlation function can be used to investigate the
later effects. Understanding the emission time sequence of
neutrons and protons will be helpful  to understand the nuclear
interaction. It also might be sensitive to the nuclear equation of
state.  It has been shown that  the emission times of the nucleons
are related to their kinetic energies. Generally, earlier emitted
nucleons have higher energies than later ones. Some results have
been demonstrated in  the experiments \cite{Ghetti2}.

\begin{figure}
\vspace{-0.2truein}
\includegraphics[scale=0.4]{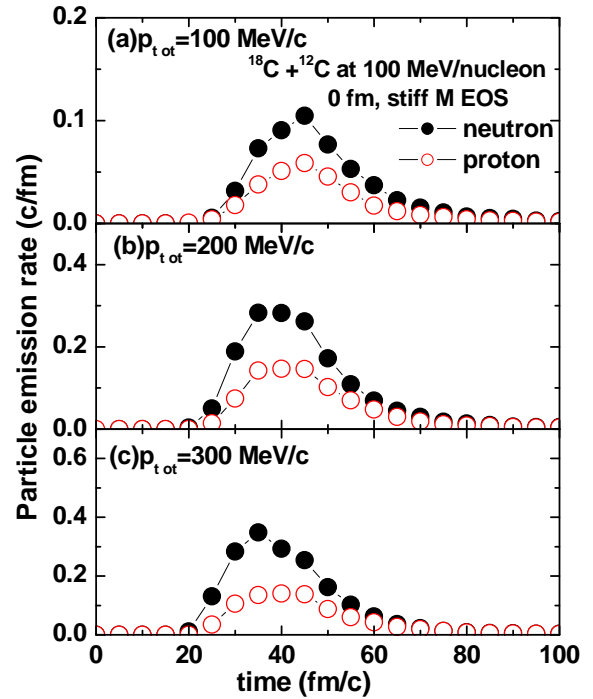}
\vspace{-0.2truein} \caption{\footnotesize (Color online) The time
evolution of the emission rate of nucleons for $^{18}$C + C at 100
MeV/nucleon and {\it b} = 0 fm with the different gates of  the
total nucleon-nucleon  pair momentum. } \label{Fig-emssiontime}
\end{figure}

Fig.~\ref{Fig-emssiontime} shows the time evolution of the
particle emission rate by using the stiff momentum-dependent
(Stiff M)  EOS. The open circles connected with solid line show
the proton emission rate and the filled ones represent the neutron
emission rate. The collisions are simulated for $^{18}$C +
$^{12}$C at 100 MeV/nucleon and {\it b} = 0 fm. In the following
discussions, except the special note, the collisions are all
performed in the above circumstance.

From this figure, nucleons begin to be emitted around 20 fm/${c}$
and their emission rates reach to their maximum 15-20 fm/${c}$
later. With increasing total nucleon-nucleon momentum $P_{\rm
tot}$,
 peak of the emission rate of nucleons becomes larger and
its corresponding time  tends to earlier time. This indicates that
nucleons with higher total pair momentum are emitted earlier.
Higher momentum nucleons mostly belong to pre-equilibrium emission
nucleons and essentially originate from higher density regions. In
contrast, lower momentum nucleons are mostly emitted from
equilibrium-like sources. For neutron-proton pairs with lower
total momentum, the emission rate of neutron and of proton is
almost synchronous, ie. there is no obvious difference of the
emission sequence between neutron and proton. However, for the
neutron-proton pair with larger total momentum, the emission rate
of neutrons reaches to the peak value earlier than that of
protons, it means that, on average, neutrons are emitted earlier
than protons. The reason why the emission rate of neutrons is
larger than that of protons stems from the neutron-rich content of
projectile. Two interpretations seem to be possible. On  one hand,
the symmetry potential term in Eq.(2) plays an important role in
controlling the emission of nucleons. In neutron-rich projectile
induced reaction,  protons could feel stronger attractive
potential due to neighboring neutrons which results in more bound
protons for disassembling sources. Viceversa, neutrons will, on
one side, feel the stronger repulsive interaction due to more
neutron-neutron pairs and, on the other side, feel smaller
attractive potential because of the decreasing of the average
assortative number of nearest neighboring protons for a certain
neutron for increasing isospin of the source. Both reasons will
lead to produce more unbound neutrons for disassembling sources
with higher isospin \cite{Ma_PRC-iso}.

\begin{figure}
\vspace{-0.2truein}
\includegraphics[scale=0.4]{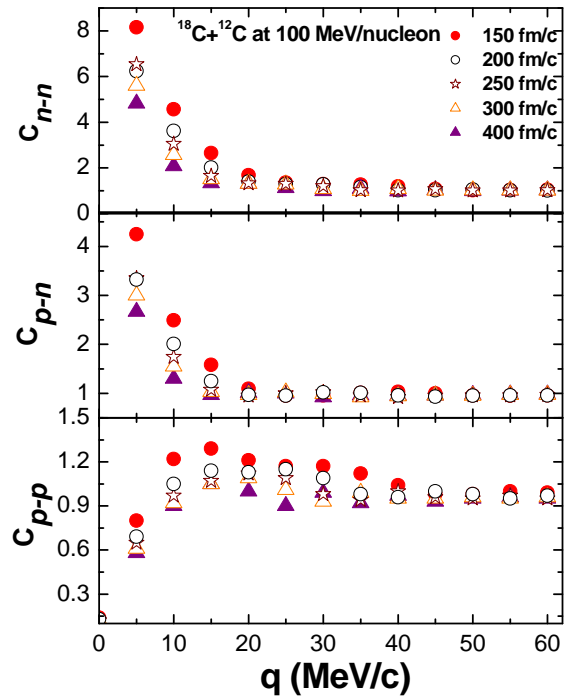}
\vspace{-0.2truein} \caption{\footnotesize (Color online) The
momentum correlation functions of n-n, p-n and p-p pairs for
$^{18}$C + C at 100 MeV/nucleon and {\it b} = 0 fm are constructed
in different evolution  time of the system. The evolution time is
illustrated in the insert. } \label{Fig_Tdep}
\end{figure}

Experimentally, the momentum correlation function reflects the
information in the final state of the reaction. Theoretically, the
final state can be seen as the state at the freeze-out of system.
Since the HBT is sensitive to the space-time information, we shall
investigate the HBT at different evolution times of reaction.
Fig.~\ref{Fig_Tdep} shows the correlation functions of
neutron-neutron, proton-neutron and proton-proton when the
evolution time of reaction $t$ = 150, 200, 250, 300 and 400
fm/${c}$. Generally, HBT values become smaller in later evolution
time due to the weakness of the nucleon-nucleon correlation when
the system is diluted. However, we can roughly say that the HBT
values do not change dramatically after $t$ = 200 fm/${c}$ after
$t$ = 200 fm/${c}$ is compared to earlier times. Because of the
limited computation resource, thereafter we investigate the
features of HBT for all systems when $t$ = 200 fm/${c}$.

\subsection{ Gate of the total momentum}

\begin{figure}
\vspace{-0.4truein}
\includegraphics[scale=0.4]{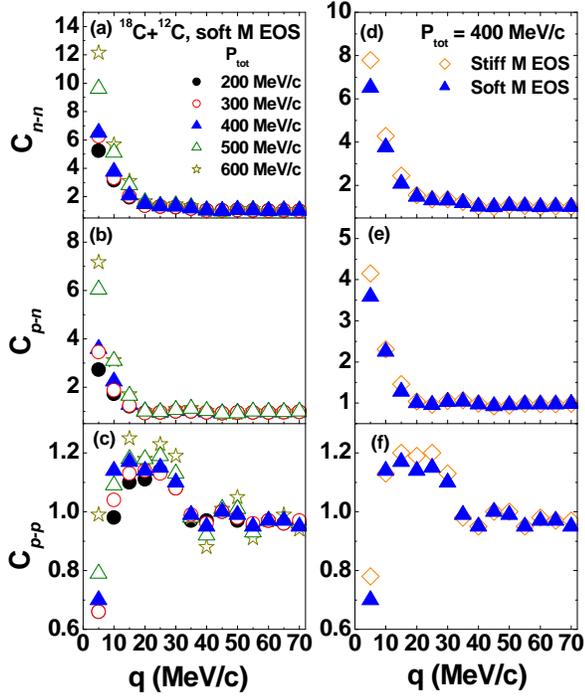}
\vspace{-0.truein} \caption{\footnotesize (Color online) The
two-nucleon correlation function $C_{\it n-n}$, $C_{\it p-n}$ and
$C_{\it p-p}$ for $^{18}$C + C at 100 MeV/nucleon and {\it b} = 0
fm with the different total pair momentum ($P_{\rm tot}$) cuts or
EOS parameters: the left panels correspond to the different gates
of $P_{\rm tot}$ while the right panels correspond to different
EOS for $P_{\rm tot}$ = 400 MeV/{\it c}. }
\label{Fig-P-gate-shape}
\end{figure}

Since the magnitude  of the total pair momentum is related to the
nucleon emission time, we shall discuss the effect of the total
pair momentum on HBT in this section. Earlier emission time
induces stronger correlation,  larger total momentum thus
contributes to the strength of correlation function too. We
discuss the calculations with  different total nucleon-nucleon
pair momentum ( $P_{\rm tot}$ ) in the following. The results are
shown in Fig.~\ref{Fig-P-gate-shape}. In this figure, three types
of nucleon-nucleon correlation function, namely that of n-n, n-p
and p-p, are shown. From the left panels, it is cleanly observed
that the higher the $P_{\rm tot}$, the higher the HBT value in
lower relative momentum. While, from the right panels, it is
observed that HBT value in the region of lower relative momentum
is higher with the soft EOS than that with the stiff EOS.

\begin{figure}
\vspace{-0.4truein}
\includegraphics[scale=0.3]{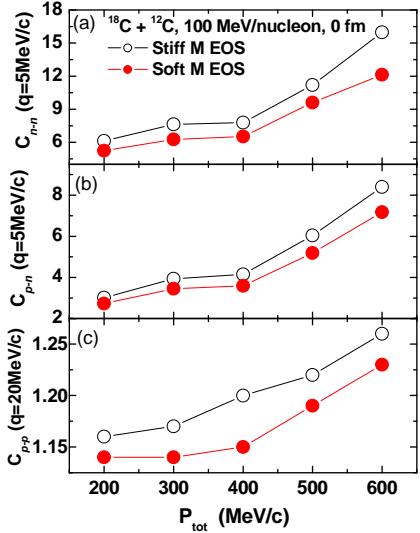}
\vspace{-0.truein} \caption{\footnotesize (Color online) The
 HBT strength of two-nucleon correlation
function for n-n  and p-n at 5 MeV/{\it c} or at p-p at 20
MeV/{\it c} for $^{18}$C + C at 100 MeV/nucleon as a function of
the gate of the total momentum ($P_{\rm tot}$) of the
nucleon-nucleon pairs in the collisions. } \label{Fig-P-gate}
\end{figure}

To see an overall trend of strength of HBT versus $P_{\rm tot}$,
we plot the value of $C_{\it n-n}$ ($C_{\it p-n}$) at 5 MeV/{\it
c} or $C_{\it p-p}$ at 20 MeV/{\it c} as a function of $P_{\rm
tot}$ in Fig.~\ref{Fig-P-gate}. The filled circles connected with
the solid line present the results with the soft
momentum-dependent EOS  and the open circles connected with the
solid line show the ones with the stiff momentum-dependent EOS.

From the figure, it is clear that the strength of two-nucleon
correlation function is smaller at lower total pair momentum than
that of the higher one. As it is shown the nucleons with lower
total momenta are emitted later than those with higher total
momenta which naturally reduces the HBT strength. The calculated
difference indicates the qualitative characterization of the
emission process during the collisions. On the other hand, the
tendency is similar despite the different combination of
nucleon-nucleon pair. Experimental results for  the momentum-gated
nucleon pairs  show a trend similar to the one that can be found
in the literature, see for example  Ref. ~\cite{Gong,Colonna}.

In addition, one can find that  the HBT strength
 with the stiff momentum-dependent EOS is higher than that with the
soft one. The influence of the different momentum dependent EOS on
HBT  will be discussed in the next section in details.

\subsection{ Soft and stiff momentum-dependent potential}

The EOS is considered an important property of nuclear matter and
several researches have been performed to investigate on the EOS
of the finite nuclear matter \cite{Lynch2}. In this section, we
will show the results of correlation functions with different EOS
parameters.

\begin{figure}
\vspace{-0.4truein}
\includegraphics[scale=0.4]{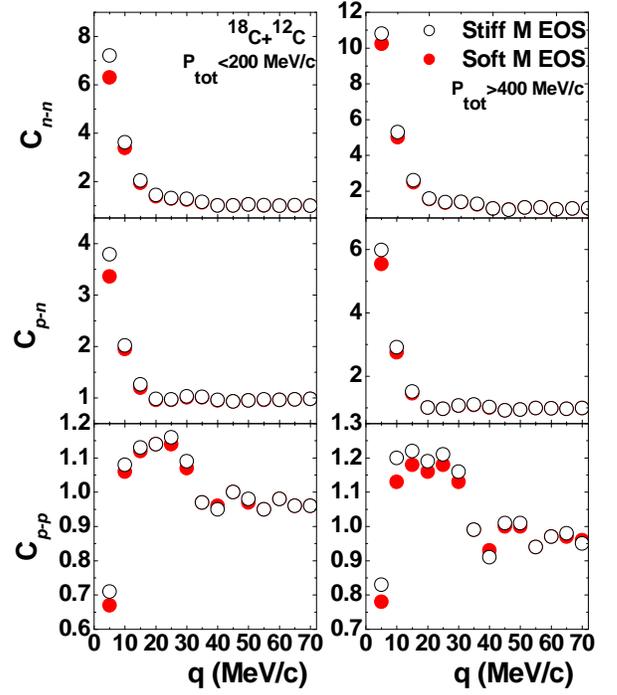}
\vspace{-0.2truein} \caption{\footnotesize (Color online) The
comparison between the correlation function  with soft
momentum-dependent (Soft M) and stiff momentum-dependent (Stiff M)
EOS for $^{18}$C + C at 100 MeV/nucleon  and {\it b} = 0fm in
IDQMD calculations. The left panels correspond to the cases of
$P_{\rm tot} <$ 200 MeV/{\it c} while the right panels correspond
to the cases of $P_{\rm tot} >$ 400 MeV/{\it c}. The filled and
open circles represent the results with Soft- and Stiff- M EOS,
respectively.} \label{Fig_eos}
\end{figure}

In previous studies, the role of  different potentials, i.e, the
soft and stiff potential in the transport models, has been
investigated via some physical observables. However, in most case,
the potential does not include the momentum dependent term, eg. in
HBT studies with isospin dependent BUU \cite{Chen1}. In this work
we used a momentum-dependent part in the potential, namely the
soft momentum-dependent potential (Soft M EOS) and the stiff
momentum-dependent potential (Stiff M EOS). we use also an
isospin-dependent potential and calculated nucleon-nucleon
correlation functions with the above mentioned two types of
momentum-dependent potentials.  The results are shown in
Fig.~\ref{Fig_eos}.

In Fig.~\ref{Fig_eos} the filled circles connected with solid line
represent the correlation function  with Soft M EOS and the open
ones are the results with the one with Stiff M EOS. The left parts
are the HBT results with $P_{\rm tot} < $ 200 MeV/{\it c} and the
right parts are the results $P_{\rm tot} > $  400 MeV/{\it c}.
From the figure, it is clear that the correlation function  with
the stiff potential is larger than that with the soft potential,
which is similar to the  calculation results with BUU \cite{Gong}.
Stiff potential makes the compression of nucleonic matter
difficult compared with the case with the soft one and leads  to
larger emission rate and earlier average emission time of
nucleons, which leads to stronger  correlation function for the
stiff potential.

\subsection{In-medium nucleon-nucleon  cross section}

\begin{figure}
\vspace{-0.4truein}
\includegraphics[scale=0.4]{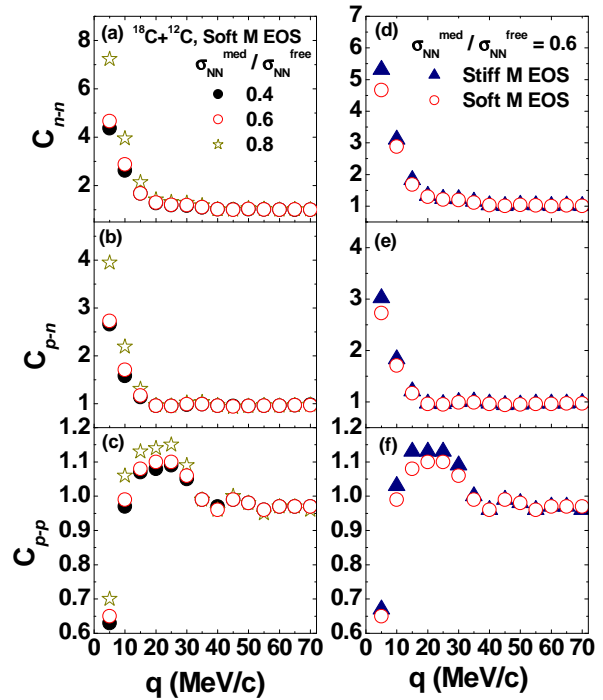}
\vspace{-0.2truein} \caption{\footnotesize (Color online)  The
two-nucleon correlation function $C_{\rm NN}$, $C_{\it p-n}$ and
$C_{\it p-p}$ for $^{18}$C + C at 100 MeV/nucleon and {\it b} = 0
fm with different in-medium nucleon-nucleon cross section
($\sigma_{\rm NN}^{\rm med}/\sigma_{\rm NN}^{\rm free}$) (left
panels) or different EOS (right panels) for fixed 0.6$\sigma_{\rm
NN}^{\rm free}$. } \label{Fig_nncrs-shape}
\end{figure}

\begin{figure}
\vspace{0.truein}
\includegraphics[scale=0.3]{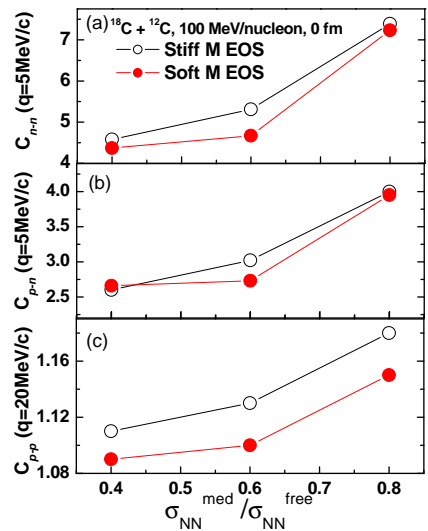}
\vspace{-0.2truein} \caption{\footnotesize (Color online)  The HBT
strengths of NN (a) and PN (b) at 5 MeV/{\it c} and PP (c) at 20
MeV/{\it c}
 as a function of the in-medium
nucleon-nucleon cross section ($\sigma_{\rm NN}^{\rm
med}/\sigma_{\rm NN}^{\rm free}$) from the reaction $^{18}$C +
$^{12}$C at 100 MeV/nucleon and {\it b} = 0 fm. The open and
filled symbols represent stiff and soft momentum-dependent
potential, respectively.} \label{Fig_nncrs}
\end{figure}

The effect of the in-medium nucleon-nucleon cross section
($\sigma_{\rm NN}^{\rm med}$) is discussed in this section. We use
a value of $\sigma_{\rm NN}^{\rm med}$ which is different from the
free nucleon-nucleon cross section ($\sigma_{\rm NN}^{\rm free}$)
to investigate its influence on momentum correlation functions.
Fig.~\ref{Fig_nncrs-shape} shows NN, NP and PP correlation
functions for different  $\sigma_{\rm NN}^{\rm med}$ and EOS.
Slightly larger values of the HBT is found for larger $\sigma_{\rm
NN}^{\rm med}$, especially for 0.8 $\sigma_{\rm NN}^{\rm free}$,
as well as for the stiff EOS. Fig.~\ref{Fig_nncrs} shows the HBT
strengths at 5 MeV/{\it c} (for NN and NP) or at 20 MeV/{\it c}
(for PP) as a function of $\sigma_{\rm NN}^{\rm med}$ in the stiff
and soft EOS. The strength increases with  $\sigma_{\rm NN}^{\rm
med}$ and with the stiffness of the EOS. This can be understood by
the following arguments: with the increasing of the in-medium
nucleon-nucleon cross section, the collision rate between two
nucleons increases and consequently the system reaches to the
equilibrium stage faster. Before equilibrium is reached, more
pre-equilibrium nucleons are emitted, which makes the strength of
the correlation function larger.

\subsection{ Impact parameter dependence}

\begin{figure}
\vspace{-0.2truein}
\includegraphics[scale=0.4]{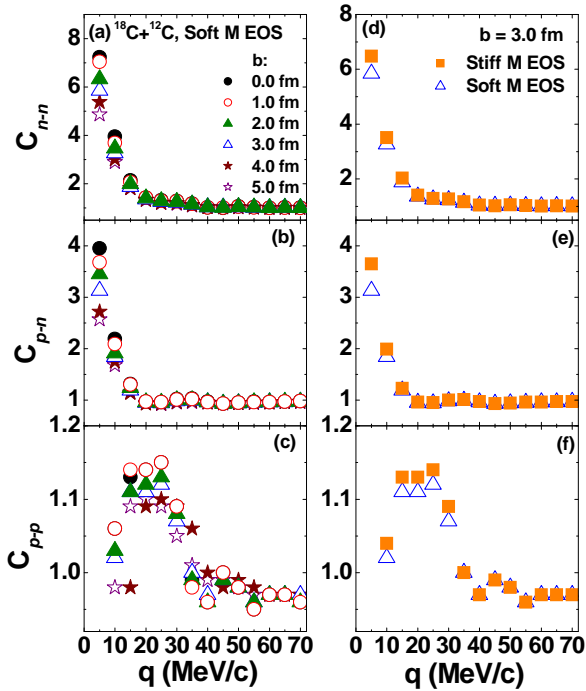}
\vspace{-0.2truein} \caption{\footnotesize (Color online) Momentum
correlation functions for n-n, p-n and p-p are calculated in
different impact parameters (left
 panels) or with different EOS at fixed {\it b} = 3 fm for $^{18}$C + C at 100 MeV/nucleon.  }
\label{Fig_Bdep-shape}
\end{figure}

\begin{figure}
\vspace{-0.2truein}
\includegraphics[scale=0.3]{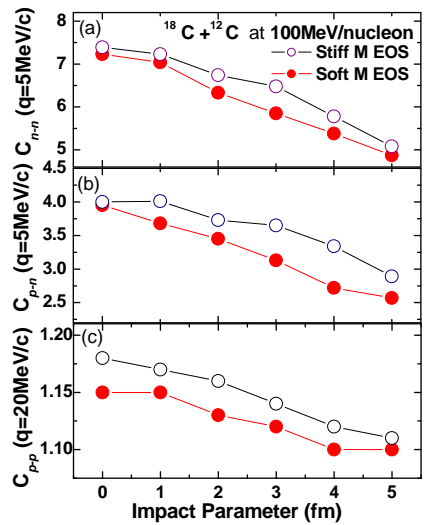}
\vspace{-0.1truein} \caption{\footnotesize (Color online) The
strength of the nucleon-nucleon correlation functions of  n-n (a)
and NP (b) at 5 MeV/{\it c}  or PP (C) at 20 MeV/{\it c} for
$^{18}$C + C at 100 MeV/nucleon as a function of  impact
parameter. The open and filled  symbols represent stiff and soft
momentum-dependent potential, respectively.} \label{Fig_Bdep}
\end{figure}

Considering the importance of the Wigner function which depends on
the impact parameter, we shall investigate on momentum correlation
function at different impact parameter. Collisions of $^{18}$C +
$^{12}$C  are performed at 100 MeV/nucleon and at the impact
parameters of 0, 1, 2, 3, 4 and 5 fm. The calculated total
momentum integrated correlation functions are shown in
Fig.~\ref{Fig_Bdep-shape}. Larger HBT values for central
collisions or stiff EOS are predicted. Their strengths are shown
in Fig.~\ref{Fig_Bdep}.

In Ref.~\cite{Gong}, some explanations on the effect of impact
parameter have been presented. As we already know, the strength of
the  correlation function mainly depends on the emission time and
the source size.  From the figure,  there exists a large
difference between soft and stiff potentials. Secondly, the
strength of correlation function becomes weaker with increasing
impact parameter on both soft and stiff momentum-dependent EOS.
This indicates that the stiffness of the potential in the IDQMD
does not change the tendency of the HBT strength of HBT with
increasing impact parameter. On the other hand, the behavior of
the HBT strength with impact parameter might reflect the changed
size of the emitting source. In central collisions,
nucleon-nucleon collisions are very frequent and emitted nucleons
are mostly from one compact and hot dense region. Therefore, the
HBT strength is larger because of the smaller source size, if
compared to peripheral collisions. Also some other reasons
including Fermi jets may contribute to this effect \cite{Fuchs}.

\subsection{Incident energy dependence}

\begin{figure}
\vspace{-0.2truein}
\includegraphics[scale=0.40]{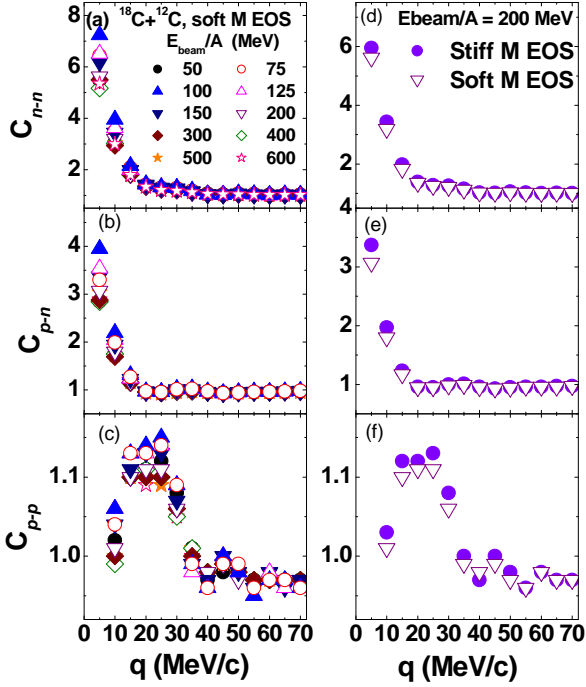}
\vspace{-0.2truein} \caption{\footnotesize (Color online) The
momentum correlation function of n-n (a), p-n (b) and p-p (c)
pairs in different beam energies in head-on collisions for soft M
EOS (left panels) or different EOS at $E_{\rm beam}$ = 200
MeV/nucleon (right panels). } \label{Fig_Edep-shape}
\end{figure}

\begin{figure}
\vspace{-0.2truein}
\includegraphics[scale=0.35]{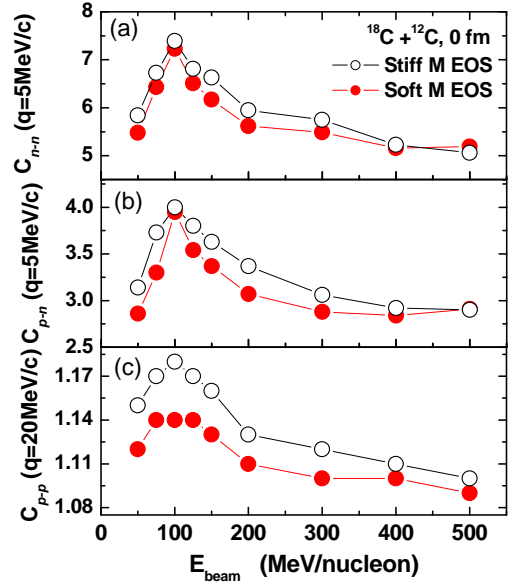}
\vspace{-0.2truein} \caption{\footnotesize (Color online) HBT
strengths of n-n (a), p-n (b) and p-p (c) as a function of beam
energy in head-on collisions. Open or filled circles represent the
calculations with stiff or soft momentum potential, respectively.}
\label{Fig_Edep}
\end{figure}

The influence of the incident energy on HBT is  investigated in
this section. Fig.~\ref{Fig_Edep-shape} shows the momentum
correlation function at different beam energies (left panels) and
different EOS (right panels). Fig.~\ref{Fig_Edep} displays the
calculated HBT strengths of n-n (a), p-n (b) and p-p (c) as a
function of beam energy for head-on collisions. From the figure,
both stiff and soft momentum potential provides a similar
evolution of the HBT strength with beam energy.  Interestingly,
the HBT strength first increases with the incident energy and
reaches a peak around 100 MeV/nucleon and then decreases at higher
incident energies. The raise of the HBT strength at lower incident
energies can be essentially attributed to drastic nucleon-nucleon
collision at higher energies, which results, on average, in
earlier emission of nucleons. However, the decreasing behavior of
HBT strengths with incident energies higher than 100 MeV/nucleon
cannot be explained in this way.

\begin{figure}
\vspace{-0.5truein}
\includegraphics[scale=0.35]{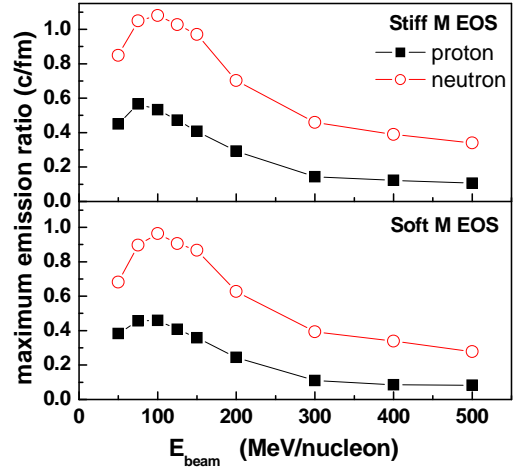}
\vspace{-0.6truein} \caption{\footnotesize (Color online) The
maximum nucleon emission rate as a function of the incident
energy. Squares and circles represent proton and neutron,
respectively. } \label{Fig_emis-Edep}
\end{figure}

In order to discuss the possible origin of the  complex behavior
of the HBT strength with incident energy, we investigate   the
maximum  nucleon emission rate in the whole  time evolution as a
function of beam energy in Fig.~\ref{Fig_emis-Edep}. From the
figure, the maximum nucleon emission rates show a rise and fall
behavior similar to the one observed in Fig.~\ref{Fig_emis-Edep}
for the HBT strength. Around 100 MeV/nucleon, the nucleon emission
rate is maximum. This energy coincides with the peak position of
the HBT strength in Fig.~\ref{Fig_emis-Edep}. From this point of
view, the HBT strength could be correlated to the maximum nucleon
emission rate during heavy ion collisions. The higher the nucleon
emission rate, the stronger the HBT strength. Of course, another
possible explanation for the falling branch of HBT strength can be
attributed to the source size in the final state ($t$ = 200
fm/${c}$). It is expected that in the energy range of a few
hundreds of MeV/nucleon, where the repulsive nucleon-nucleon
interaction is dominant, a more dilute system might develop after
freeze-out. So the fall of HBT with beam energy above 100
MeV/nucleon indicates that the source size has been much expanded
in the final state at higher energies. In contrast, the attractive
mean field competes with the repulsive nucleon-nucleon interaction
below 100 MeV/nucleon. Shorter emission times with increasing beam
energies may play a dominant role in determining the behavior of
the HBT strength. Fig.~\ref{Fig_emis-Edep} also illustrates
illustrates that the emission rate of neutrons is larger than that
of protons and that the neutrons are emitted earlier than protons.

 \section{Summary}

In summary, isospin-dependent quantum dynamics model has been used
as an event-generator to study the momentum correlation function
of neutron-neutron, proton-proton and neutron-proton correlation
functions for C + $^{12}$C. An approximate linear relationship
between the strength of correlation function at low relative
momentum and the mean binding energy of projectiles has been
revealed and the influence of the different initialization methods
of the projectiles on momentum correlation function has been also
investigated. In addition, some different physical ingredients
which have effects on momentum correlation functions have been
shown in this work. These ingredients include nuclear equation of
state and in-medium nucleon-nucleon cross section. It shows that
the stiff momentum-dependent EOS or a larger in-medium
nucleon-nucleon cross section results in stronger HBT strength
than the soft momentum-dependent EOS or a smaller in-medium
nucleon-nucleon cross section. Some other aspects which have also
effects on momentum correlation function,  including the time
evolution of the reaction system, the impact parameter, the gate
of the total momenta of nucleon-nucleon pairs and the incident
energy are also explored. Results show that the HBT strength
decreases  with increasing impact parameter due to the earlier
emission or a compact source size in central collisions. A
positive correlation between the gate of total nucleon-nucleon
pair momentum and the HBT strength can be explained by the earlier
emission time of nucleons with higher total momentum. The rise and
fall of the HBT strengths with the beam energy can be interpreted
either in terms of the maximum nucleon emission rates during
heavy-ion collisions or in terms of  the dominant shorter emission
time with increasing beam energy  below 100 MeV/nucleon  and the
larger later-stage expansion of the source size with the
increasing of beam energy above 100 MeV/nucleon.

\section*{Acknowledgments}
This work was supported in part by the Shanghai Development
Foundation for Science and Technology under Grant Numbers
05XD14021 and 03 QA 14066, the  National Natural Science
Foundation of China  under Grant No 10328259, 10135030, 10535010
the Major State Basic Research Development Program under
 Contract No G200077404.

{}

\footnotesize
\begin{thebibliography}{}

\bibitem{Brown} R. Hanbury Brown, R.Q. Twiss, Nature {\bf 178},   1046 (1956).


\bibitem{Goldhaber} G. Goldhaber, S. Goldhaber, W. Lee, and A. Pais, Phys. Rev. {\bf
120}, 300 (1960).

\bibitem{Heinz2} U. Heinz and B. Jacak, Ann. Rev. Nucl. Part. Sci. {\bf 49}, 529
(1999).


\bibitem{Boal} D. H. Boal, C. K. Gelbke, B. K. Jennings, Rev. Mod. Phys. {\bf 62}, 553
(1990).

\bibitem{Bauer} W. Bauer,  C. K. Gelbke, S. Pratt, Ann. Rev. Nucl.
Part. Sci. {\bf 42}, 77 (1992).

 \bibitem{Wiedman}U. A. Wiedemann and U. Heinz, Phys. Rep. {\bf 319}, 145 (1999).

\bibitem{Henny} M. Henny, S. Oberholzer, C. Strunk, T. Heinzel, K. Ensslin, M. Holland, and C.
Sch\"onenberger, Science {\bf 284}, 296 (1999); William D. Oliver,
Jungsang Kim, Robert C. Liu, and Yoshihisa Yamamoto,  Science {\bf
284}, 299 (1999).

\bibitem{Picciotto} R. de-Picciotto, M. Reznikov, M. Heiblum, V. Umansky, G. Bunin,
D. Mahalu, Nature {\bf 389}, 162 (1997).

\bibitem{Lynch} W. G. Lynch, C. B. Chitwood, M. B.Tsang, D. J. Fields,
D. R. Klesch, C. K. Gelbke, G. R. Young, T. C. Awes, R. L.
Ferguson, F. E. Obenshain, F. Plasil, R. L. Robinson, and A. D.
Panagiotou,  Phys. Rev. Lett. {\bf 51}, 1850 (1983).

\bibitem{Pochodzalla} J. Pochodzalla, C. K. Gelbke, W. G. Lynch, M. Maier,
D. Ardouin, H. Delagrange, H. Doubre, C. Gr\'egoire, A. Kyanowski,
W. Mittig, A. P\'eghaire, J. P\'eter, F. Saint-Laurent,  B.
Zwieglinski, G. Bizard, F. Lef\'ebvres, B. Tamain, J. Qu¨¦bert, Y.
P. Viyogi, W. A. Friedman, D. H. Boal, Phys. Rev. C {\bf 35}, 1695
(1987).

\bibitem{Gong} W. G. Gong, W. Bauer, C.K. Gelbke, S. Pratt, Phys. Rev. C {\bf 43},  781 (1991).

\bibitem{Colonna} N. Colonna, D.R. Bowman, L. Celano,
  G. D'Erasmo, E. M. Fiore, L. Fiore, A. Pantaleo, V. Paticchio,
  G. Tagliente, and S. Pratt, Phys. Rev. Lett. {\bf 75},  4190 (1995).



\bibitem{Ghetti} R. Ghetti, V. Avdeichikov, B. Jakobsson,
  P. Golubev, J. Helgesson, N. Colonna, G. Tagliente, H.W. Wilschut,
  S. Kopecky, V.L. Kravchuk, E.W. Anderson, P. Nadel-Turonski,
  L. Westerberg, V. Bellini, M.L. Sperduto, C. Sutera, Phys.Rev. C {\bf 69},  031605
  (2004).


\bibitem{Orr} N. A. Orr, Nucl. Phys. A {\bf 616}, 155 (1997).

\bibitem{Marques1} F. M. Marques, M. Labiche, N.A. Orr,
J. C. Ang\'elique, L. Axelsson, B. Benoit, U. C. Bergmann, M. J.
G. Borge, W. N. Catford, S. P. G. Chappell, N. M. Clarke, G.
Costa, N. Curtis, A. D'Arrigo, F. de Oliveira Santos, E. de G\'oes
Brennand, O. Dorvaux, M. Freer, B. R. Fulton, G. Giardina, C.
Gregorie, S. Gr\'evy, D. Guillemaud-Mueller, F. Hanappe, B.
Heusch, B. Jonson, C. Le Brun, S. Leenhardt, M. Lewitowicz, M. J.
L\'opez, K. Markenroth, M. Motta, A. C. Mueller, T. Nilsson, A.
Ninane, G. Nyman, I. Piqueras, K. Riisager, M. G. Saint Laurent,
F. Sarazin, S. M. Singer, O. Sorlin and L. Stuttg\'e, Phys. Lett.
B {\bf 476}, 219 (2000).

\bibitem{Marques2} F. M. Marques, M. Labiche, N.A. Orr,
J. C. Ang\'elique, L. Axelsson, B. Benoit, U. C. Bergmann, M. J.
G. Borge, W. N. Catford, S. P. G. Chappell, N. M. Clarke, G.
Costa, N. Curtis, A. D¡¯Arrigo, E. de G\'oes Brennand, F. de
Oliveira Santos, O. Dorvaux, G. Fazio, M. Freer, B. R. Fulton, G.
Giardina, S. Gr\'evy, D. Guillemaud-Mueller, F. Hanappe, B.
Heusch, B. Jonson, C. Le Brun, S. Leenhardt, M. Lewitowicz, M. J.
L\'opez, K. Markenroth, A. C. Mueller, T. Nilsson, A. Ninane, G.
Nyman, I. Piqueras, K. Riisager, M. G. Saint Laurent, F. Sarazin,
S. M. Singer, O. Sorlin, and L. Stuttg\'e,  Phys. Rev. C {\bf 64},
061301 (2001).

\bibitem{Verde1}G. Verde, D. A. Brown, P. Danielewicz, C. K. Gelbke, W. G. Lynch, and
 M. B. Tsang, Phys. Rev. C {\bf 65}, 54609 (2002).

\bibitem{Verde2}G. Verde,  P. Danielewicz, D. A. Brown, W. G. Lynch,  C. K. Gelbke and
 M. B. Tsang, Phys. Rev. C {\bf 67}, 34606 (2003).

\bibitem{Buu1}D. O. Handzy, W. Bauer, F. C. Daffin, S. J. Gaff, C. K. Gelbke, T.
Glasmacher, E. Gualtieri, S. Hannuschke, M. J. Huang, G. J. Kunde,
R. Lacey, T. Li, M. A. Lisa, W. J. Llope, W. G. Lynch, L. Martin,
C. P. Montoya, R. Pak, G. F. Peaslee, S. Pratt, C. Schwarz, N.
Stone, M. B. Tsang, A. M. Vander Molen, G. D. Westfall, J. Yee,
and S. J. Yennello, Phys. Rev. Lett. {\bf 75}, 2916 (1995).


\bibitem{Bertsch}G. F. Bertsch and S. Das Gupta, Phys.\ Rep.\
{\bf 160}, 189 (1988).

\bibitem{Aichelin} J. Aichelin, Phys. Rep. {\bf 202}, 233 (1991).

\bibitem{Chen1} L. W. Chen, V. Greco, C. M. Ko, and B. A. Li,
Phys. Rev. C {\bf 68}, 014605 (2003).

\bibitem{Wei1} Y. B. Wei, Y. G. Ma, W. Q. Shen, G. L. Ma, K. Wang, X. Z. Cai,
C. Zhong, W. Guo, J. G. Chen, Phys. Lett. B {\bf 586}, 225 (2004).

\bibitem{Wei2} Y. B. Wei, Y. G. Ma, W. Q. Shen,
G. L. Ma, K. Wang, X. Z. Cai, C. Zhong, W. Guo, J. G. Chen, D. Q.
Fang, W. D. Tian, X. F. Zhou, J. Phys. G {\bf 30}, 2019 (2004).




\bibitem{Koonin} S.E. Koonin, Phys. Lett. B {\bf 70}, 43 (1977).

\bibitem{Pratt1} S. Pratt, Phys. Rev. Lett {\bf 53},  1219 (1984).

\bibitem{Pratt2} S. Pratt, M. B. Tsang,  Phys. Rev. C {\bf 36}, 2390 (1987).


\bibitem{crab} S. Pratt,   Nucl. Phys. {\bf A 566}, 103c (1994).

\bibitem{Aich2} J. Aichelin, A. Rosenhauer, G. Peilert, H. St\"ocker
and W. Greiner, Phys. Rev. Lett. {\bf 58}, 1926 (1987).

\bibitem{Westfall} G. F. Westfall, W. Bauer, D. Craig,
 M. Cronqvist, E. Gaultieri, S. Hannuschke, D. Klakow, T. Li,
 T. Reposeur, A. M. Vander Molen, W. K. Wilson, J. S. Winfield,
 J. Yee, S. J. Yennello, R. Lacey, A. Elmaani, J. Lauret,
 A. Nadasen, E. Norbeck, Phys. Rev. Lett. {\bf 71}, 1986 (1993).

\bibitem{Klakow} D. Klakow, G. Welke, and W. Bauer, Phys. Rev. C {\bf 48}, 1982
(1993).

\bibitem{Ma_flow}Y. G. Ma, W. Q. Shen, J. Feng, Y. Q. Ma, Phys. Rev. C {\bf 48}, R1492 (1993);
Y. G. Ma and W. Q. Shen, Phys. Rev. C {\bf 51}, 3256 (1995).

\bibitem{Chen3} K. Chen, Z. Fraenkel, G. Friedlander, J. R. Grover, J. M. Miller,
and Y. Shimamoto, Phys. Rev. {\bf 166}, 949 (1968).

\bibitem{Ghett_mult}R. Ghetti, J. Helgesson, N. Colonna, B. Jakobsson,
A. Anzalone, V. Bellini, L. Carl\'en, S. Cavallaro, L. Celano, E.
De Filippo, G. D¡¯Erasmo, D. Di Santo, E. M. Fiore, A. Fokin, M.
Geraci, F. Giustolisi, A. Kuznetsov, G. Lanzan\'o, D. Mahboub, S.
Marrone, F. Merchez, J. Martensson, F. Palazzolo, M. Palomba, A.
Pantaleo, V. Paticchio, G. Riera, M. L. Sperduto, C. Sutera, G.
Tagliente, M. Urrata, and L. Westerberg, Phys. Rev. C {\bf 64},
017602 (2001).

\bibitem{Ghetti2}R. Ghetti, J. Helgeson, V. Avdeichikov, P. Golubev, B. Jakobsson,
N. Colonna, G. Tagliente, S. Kopecky, V. L. Kravchuk, H. W.
Wilschut, E. W. Anderson, P. Nadel-Turonski, L. Westerberg, V.
Bellini, M. L. Sperduto, and C. Sutera,  Phys. Rev. Lett. {\bf
91},  092701 (2003).

\bibitem{Ma_PRC-iso} Y. G. Ma,  Q. M. Su, W. Q. Shen,  D. D. Han,
J. S.  Wang,  X. Z. Cai,  D. Q. Fang,  H. Y. Zhang, Phys. Rev. C
{\bf 60}, 24607 (1999).

\bibitem{Lynch2}P. Danielewicz, R. Lacey and W. G. Lynch, Science {\bf 298}, 1592
(2002).

\bibitem{Fuchs}H. Fuchs and K. Mohing, Rep. Prog. Phys. {\bf 57},
231 ( 1994) and refernces therein.

\end{thebibliography}
\end{document}